\documentclass[letterpaper]{article} 
\usepackage{aaai2026}  
\usepackage{times}  
\usepackage{helvet}  
\usepackage{courier}  
\usepackage[hyphens]{url}  
\usepackage{graphicx} 
\usepackage{natbib}  
\usepackage{caption} 
\usepackage{algorithm}
\usepackage{algorithmic}

\usepackage{amsmath,amsfonts} 
\usepackage{mathrsfs}  

\usepackage{tabularx}   
\usepackage{booktabs}   
\usepackage{multirow}   
\usepackage{makecell}   
\usepackage{array}          
\newcolumntype{Z}{>{\centering\arraybackslash}X} 
\usepackage{siunitx}        
\usepackage{subfig}    
\usepackage{enumitem}

\usepackage{float} 
\usepackage{etoolbox} 
\usepackage{threeparttable}

\usepackage{newfloat}
\usepackage{listings}
\DeclareCaptionStyle{ruled}{labelfont=normalfont,labelsep=colon,strut=off} 
\floatstyle{ruled}
\newfloat{listing}{tb}{lst}{}
\floatname{listing}{Listing}
\nocopyright 

\title{Influence Maximization in Multi-layer Social Networks \\ Based on Differentiated Graph Embeddings}
\author{
    Ronghua Lin\textsuperscript{\rm 1},
    Runbin Yao\textsuperscript{\rm 1},
    Yijia Wang\textsuperscript{\rm 1},
    Junjie Lin\textsuperscript{\rm 1},
    Zhengyang Wu\textsuperscript{\rm 1}\thanks{Zhengyang Wu is the corresponding author.},
    Yong Tang\textsuperscript{\rm 1}\textsuperscript{\rm 2}
}
\affiliations{

    \textsuperscript{\rm 1}School of Computer Science, South China Normal University, Guangzhou 510631, China \\
    \textsuperscript{\rm 2}Institute of Data Intelligence, Guangdong University of Science and Technology, Dongguan, Guangdong 523083, China \\
    \{rhlin, rbyao, wangyj, jjlin, wuzhengyang, ytang\}@m.scnu.edu.cn
}

\usepackage{bibentry}

\begin{document}

\maketitle

\begin{abstract}
Identifying influential nodes is crucial in social network analysis.
Existing methods often neglect local opinion leader tendencies, resulting in overlapping influence ranges for seed nodes. 
Furthermore, approaches based on vanilla graph neural networks (GNNs) struggle to effectively aggregate influence characteristics during message passing, particularly with varying influence intensities.
Current techniques also fail to adequately address the multi-layer nature of social networks and node heterogeneity.
To address these issues, this paper proposes Inf-MDE, a novel multi-layer influence maximization method leveraging differentiated graph embedding.
Inf-MDE models social relationships using a multi-layer network structure.
The model extracts a self-influence propagation subgraph to eliminate the representation bias between node embeddings and propagation dynamics.
Additionally, Inf-MDE incorporates an adaptive local influence aggregation mechanism within its GNN design.
This mechanism dynamically adjusts influence feature aggregation during message passing based on local context and influence intensity, enabling it to effectively capture both inter-layer propagation heterogeneity and intra-layer diffusion dynamics.
Extensive experiments across four distinct multi-layer social network datasets demonstrate that Inf-MDE significantly outperforms state-of-the-art methods.
\end{abstract}


\section{Introduction}
\label{sec:intro}
\begin{figure}[t]
  \centering
  \includegraphics[width=0.45\textwidth]{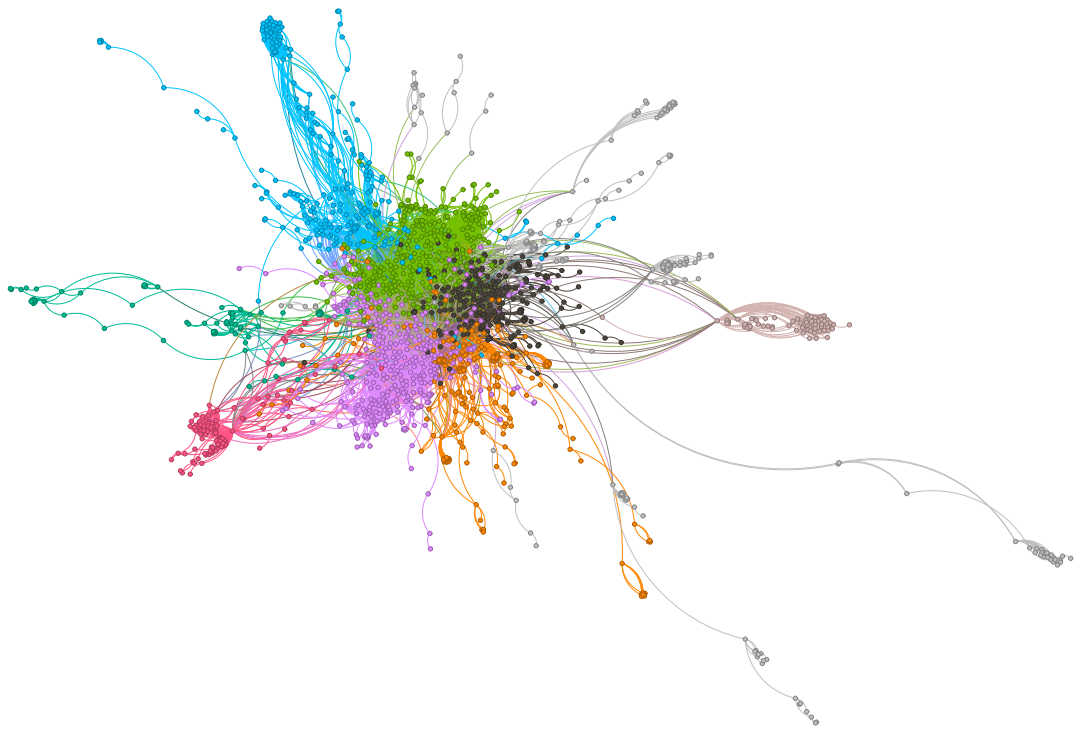}
  \caption{Visualization of multi-layer relationships in a real-world social network.}
  \label{fig:ssn}
\end{figure}

In social networks (SNs), the advent of influencer marketing has driven a class of algorithmic problems known as Influence Maximization (IM).
However, SNs are ubiquitous in the real world, a single-level modeling approach cannot reflect the complexity or distinguish between diverse relationships to describe the overall structure and interactions in social networks \cite{2023More}. 
Figure~\ref{fig:ssn} shows a real social network of scholars, where each color represents a nodal relationship, illustrating that these relationships are diverse and complex.
Relying solely on a single type of edge to represent the diverse relationships in real-world social networks often fails to capture the heterogeneity and complexity of interactions, as well as the influence propagation across distinct relationship types.

To address this limitation, multi-layer social network-based influence maximization is proposed, where the networks are characterized by homogeneous node types but heterogeneous relationship modeling.
Most current studies on the influence maximization focus on networks with single relationship type, including machine learning-based methods \cite{1219-3,1219-2}.
However, these methods are not directly applicable to multi-layer social networks as they only account for a single information propagation mechanism.
In multi-layer social networks, each layer may employ distinct influence propagation rules, leading to propagation diversity and complexity that exceeds the capabilities of conventional single-layer approaches.

Some methods extend single-layer network algorithms by aggregating layer-wise recognition results as multi-layer solutions \cite{114-4}.
However, this approach fails to capture inter-layer dependencies and intra-layer structural features during influence propagation, particularly neglecting the synergy between localized influence embeddings and global impact characterizations.

To overcome these limitations, we propose Inf-MDE (Multi-layer Influence Maximization Model based on Differentiated Graph Embedding) for multi-layer social networks.
Firstly, Inf-MDE employs information diffusion models to perform a finite number of Monte Carlo simulations and learns the local influence features by a novel Graph Local Adaptive Influence Network (GLAIN).
Subsequently, Inf-MDE conducts differentiated graph embeddings by aggregating the community information and structures of nodes.
It finally reformulates the node influence calculation as a pseudo-regression problem, where influence scores are treated as continuous values instead of discrete rankings.
These scores do not directly reflect a node's global influence rank but serve as selection indicators for optimizing seed sets.

The contributions of this paper are summarized below:
\begin{itemize}[leftmargin=*]
    \item We employ a node diffusion model to sample the cross-layer subgraph and model heterogeneous relationships, capturing both intra-layer and inter-layer propagation characteristics of nodes.
    \item We introduce a pseudo-regression strategy that integrates the positive correlation between influence scores and embedding distances, effectively mitigating overlapping influence regions within the seed set.
    \item Extensive experiments on four real-world datasets demonstrate that our proposed model outperforms state-of-the-art methods.
\end{itemize}

\section{Preliminaries}

\textbf{Multi-layer Social Network Influence Maximization (MIM)} 
For a given multi-layer social network consisting of $l$ layers $\mathcal{G} = \{ \mathcal{G}_1,\mathcal{G}_2,\dots,\mathcal{G}_l\}$, where each element is a single relational graph $\mathcal{G}_i=(\mathcal{V},\mathcal{E}_i)$ and the set of nodes $\mathcal{V} = \{v_{1},v_{2},\dots,v_{n}\}$ denotes all nodes in the social network that are shared across all layers, the edge set $ \mathcal{E}_i \subseteq \mathcal{V} \times \mathcal{V}$ denotes the relationships in the $i$-th layer of the network.
The influence maximization problem in multi-layer social networks aims to select a seed set $\mathcal{S}$ of size at most $k$ from a multi-layer social network $\mathcal{G}$ to maximize the number of eventually influenced nodes $\sigma_{\mathcal{M}}(\mathcal{S})$ under a specified information diffusion model $\mathcal{M}_i$ in each layer of the network $\mathcal{G}_i$, where $\mathcal{M} = \{\mathcal{M}_1, \mathcal{M}_2, \dots, \mathcal{M}_l\}$ denotes the corresponding diffusion model of each layer in the network.
The information diffusion models mentioned here can refer to the commonly used models, such as the Multi-layer Independent Cascade (Multi-IC)~\cite{7542505}, Multi-layer Susceptible-Infected-Recovered (Multi-SIR)~\cite{9147631}, or Multi-layer Linear Threshold (Multi-LT)~\cite{5694069} models, et al.
Based on this, the problem can be formalized as the following optimization Equ. (\ref{equ:MIM}):
\begin{equation}
  \label{equ:MIM}
  \newcommand{\argmax}{\mathop{\mathit{arg} \ \text{max}}}
  \tilde{\mathcal{S}} = \argmax_{\mathcal{S} \subseteq \mathcal{V}, \ |\mathcal{S}| \leq k} \sigma_{\mathcal{M}}(\mathcal{S}),
\end{equation}
where $\sigma_{\mathcal{M}}(\mathcal{S})$ represents the total influence expectation calculated using the Monte Carlo simulations under the given diffusion model $\mathcal{M}$ across the multi-layer social network $\mathcal{G}$.


\section{The Proposed Method}

\begin{figure*}[t]
  \centering
  \includegraphics[width=0.95\textwidth]{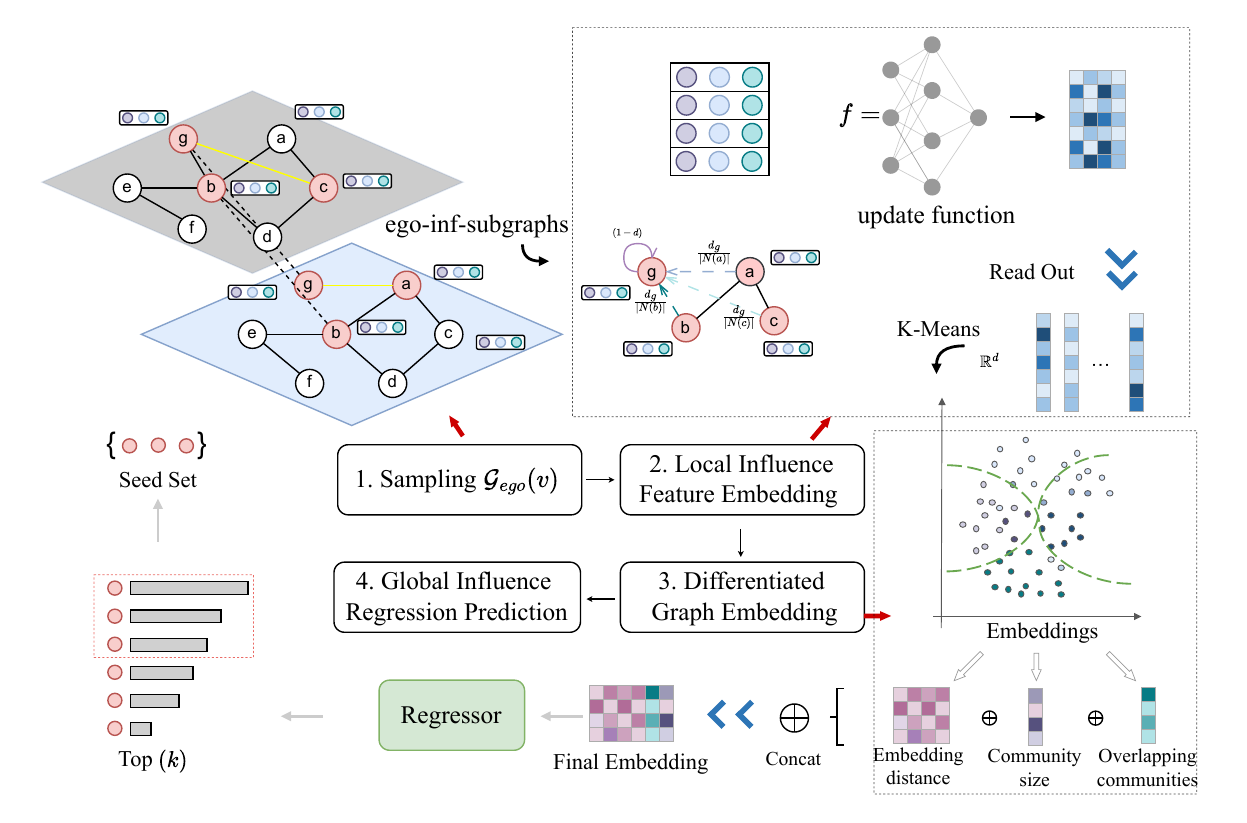}
  \caption{The architecture of our proposed model Inf-MDE.}
  \label{fig:chap4:2}
\end{figure*}

The architecture of our proposed model Inf-MDE is shown in Figure \ref{fig:chap4:2}.
It consists of four components, which are ego-inf-subgraph extraction, local influence feature embedding, differentiated graph embedding, and global influence regression prediction.

\subsection{Ego-Inf-Subgraph Extraction}
\label{subg-extra}

Most existing studies adopt breadth-first search (BFS), depth-first search (DFS), or random walks to sample the subgraphs in the influence maximization task\cite{114-6,114-8,116-4}.
However, these sampling methods generate subgraphs based solely on random paths or tree structures, failing to fully capture the topology information of influence diffusion and the hidden local influence features between nodes.

To this end, we propose a novel extraction method for the ego-center influence subgraph (called ego-inf-subgraph) sampling. 
The task is to extract the corresponding ego-inf-subgraph for each node $v$ as shown in Equ. (\ref{equ:chap4:2}).
\begin{equation}
  \label{equ:chap4:2}
  \mathcal{G}_{\text{ego-inf}} \left ( v \right ) = \left ( A(v), E_{A(v)} \right ),
\end{equation}
where $A(v)$ represents the set of nodes activated by node $v$, encompassing $v$ itself, its directly activated neighbors, and nodes subsequently activated by these neighbors, with $E_{A(v)}$ denoting the set of edges among these activated nodes, as defined in Equ. (\ref{equ:chap4:3}).
\begin{equation}
  \label{equ:chap4:3}
  E_{A(v)} = \left \{ (u, w) \in E \wedge u, w \in A(v) \right \}
\end{equation}


The pseudo-code of the proposed ego-inf-subgraph extraction method is provided in the Appendix.
Firstly, each node $ v \in \mathcal{V} $ is designated as a seed node (i.e., an active node), while all other nodes are considered inactive.
Subsequently, in each social network layer $\mathcal{G}_i$ where $i \in [1, l]$, each edge $(v,v_j)$ is assigned a predefined and constant influence probability $p^i{(v,v_j)}$.
At time step $(t-1)$ where $t \in (0, step]$, if node $v_j$ is inactive and it has multiple active neighbors, the order in which these neighbors attempt to activate $v_j$ is arbitrary.
Let $N^i_{\text{active}}(v)$ denote the set of active neighbors of node $v$ at layer $i$, the probability of its neighbor node $v_j$ being activated is as shown in Equ. (\ref{equ:chap4:4}).
\begin{equation}
  \label{equ:chap4:4}
  P_{\text{activated}}(v_j) = 1 - \prod_{u \in N^{i}_{\text{active}}(v)} \left( 1 - p^i_{(u,v_j)} \right)
\end{equation}

Furthermore, if node $v_j$ is activated from an inactive state at time step $(t-1)$, it is granted a single opportunity in the $t$-th Monte Carlo simulation to attempt activation of its inactive neighbors $N^i_{\text{un-acted}}(v_j)$ within the same social network layer.
Whether this attempt succeeds or not, $v_j$ will not retry to activate its inactive neighbors in subsequent steps.
Meanwhile, when $v_j$ is activated, it determines whether to propagate the influence information to other layers, activating the same node $v_j$ in each other layer with probability $\theta^i$.
The Monte Carlo simulation process will terminate when the number of simulation steps reaches $step$, or when no new nodes in any layer are newly activated.
The ego-inf-subgraph $\mathcal{G}_{\text{ego-inf}}(v)$ of node $v$ is constituted by its activated-node set $\mathcal{V}_{activated}(v)$ and the edges connecting them $\mathcal{E}_{acted}(v)$, where $\mathcal{G}_{\text{ego-inf}}(v)$ is originating from the ego-centric node $v$ and $\mathcal{V}_{activated}(v)$ consists of node $v$, its directly activated neighbors, and nodes subsequently activated by these neighbors.

Compared to existing subgraph sampling methods, the proposed ego-inf-subgraph extraction method minimizes the inclusion of topological structures irrelevant to information diffusion, thereby reducing uncertainty in the node embedding vector space during subsequent local influence feature generation.
To balance the scope of local influence information and computational cost, we constrain the simulation steps $step$, initial propagation probability $p^i$, and cross-layer propagation probability $\theta^i$, preventing excessively large subgraphs that could increase computational overhead in local influence embedding.
The value of $step$ will be analyzed and determined in the hyperparameter analysis.

\subsection{Local Influence Feature Embedding}


\subsubsection{Graph Embedding via GLAIN}

To address the challenge of varying cross-layer propagation probabilities among nodes, inspired by both the Graph Isomorphism Network (GIN) and the classical PageRank algorithm, we utilize a novel Graph Local Adaptive Influence Network (GLAIN) to aggregate local influence features and learn coarse-grained structural representations of ego-inf-subgraphs within multi-layer social networks.
Specifically, following the main thinking of PageRank that calculates the node influence information through iterative propagation, we propose an adaptive local influence aggregation mechanism, aiming to capture deep influence relationships between nodes by aggregating local influence features from their neighbors.
Unlike the PageRank algorithm, where nodes have a certain probability of transitioning from a random node without relying on neighbor influence, in the influence maximization problem, most information propagation occurs through links between nodes, i.e., via neighboring nodes.
The importance of aggregating these influence features is controlled by the damping factor $ d_v $, where a larger $ d_v $ indicates greater importance of neighboring nodes, causing the node’s local influence to depend more heavily on its neighbors’ influence, while a smaller $ d_v $ implies a greater reliance on the node’s own influence features.
In each round of information propagation, the features of node $ v $ are weighted and propagated based on the local influence features of its neighbors. 
The node representation update in GLAIN is shown in Equ. (\ref{equ:chap4:5}).
\begin{equation}
  \label{equ:chap4:5}
  h_v^{(t+1)} =\sigma \left( d_v \cdot \sum_{u \in \mathcal{N}(v)} \frac{h_u^{(t)}}{|\mathcal{N}(u)|} + (1 - d_v) \cdot h_v^{(t)} \right),
\end{equation}
where $\mathcal{N}(v)$ is the neighbor-node set of node $v$, $|\mathcal{N}(u)|$ denotes the size of the neighbor-node set of node $u$, $d_v$ is the damping factor for node $v$, which controls the weight of features aggregated from its neighbors, and $\sigma(\cdot)$ is a Sigmoid activation function. 

However, in Equ. (\ref{equ:chap4:5}), a fixed damping factor $d_v$ may fail to capture cross-layer relationships, particularly in the influence maximization problem, where nodes’ propagation capabilities differ across network layers.
To address this issue, we utilize a linear transformation network to adaptively learn the damping factor $ d_v $ of each node in different network layers to adjust the influence weights on neighbors’ features during information aggregation, as shown in Equ. (\ref{equ:chap4:6}).
\begin{equation}
  \label{equ:chap4:6}
  d_v = \sigma(W_d \cdot h_v + b_d),
\end{equation}
where $h_v$ is the feature embedding of node $v$ output by GLAIN, $W_d$ and $b_d$ are learnable parameters.

\subsubsection{READOUT Based on the Number of Shortest Paths}

READOUT, also called graph coarsening, is a process that aggregates the node embeddings at the last GNN layer to generate a finite-dimensional feature embedding of the entire graph \cite{121-1,121-2,121-3}, as shown in Equ. (\ref{equ:chap4:7}).
\begin{equation}
  \label{equ:chap4:7}
  h_G = \text{READOUT} \left ( \left\{ h^{(T)}_{v} \mid v \in G \right\} \right )
\end{equation}
where $h_G \in \mathbb{R}^{D}$ and $D$ is the size of dimensions, $h^{(T)}_{v}$ is the node embedding of node $v$ at the last GLAIN layer.
Since the nodes with a higher number of shortest paths passing through them contribute more significantly to the feature embeddings for influence propagation, we propose a novel READOUT function that employs a weighted summation based on the number of shortest paths to generate subgraph-level embeddings.
Specifically, as the number of shortest paths traversing a node increases, the influence of that node’s feature information on the full-graph-level embedding during the readout process correspondingly amplifies.
Thus, the READOUT function in Inf-MDE, which leverages the number of shortest paths, is formulated as shown in Equ. (\ref{equ:chap4:8}).
\begin{equation}
  \label{equ:chap4:8}
  h_G = \text{READOUT} \left( h_{v} \right) =  \sum_{v \in G}^{n} {\delta_{st}(v)} \cdot h^{(T)}_{v}
\end{equation}
where $v \in G $, $\delta_{st}(v)$ denotes the number of shortest paths from any source node $s$ to the target node $t$ that pass through node $v$.
Here, $h_G$ is the local influence feature embeddings, which are also called subgraph-level embeddings.

\subsection{Differentiated Graph Embedding}

In the influence maximization task, selecting a seed set based solely on the similarity of influence ranges and structural features is always suboptimal.
It tends to result in the seed set propagating influence to overlapping nodes, thereby reducing the overall influence coverage.
To tackle this issue, we propose a node-differentiated processing module based on community detection.

In the subgraph-level local influence embeddings $h_G \in \mathbb{R}^{D}$, the similarity between node embeddings directly reflects the similarity in their local propagation ranges and structures.
Nodes with closer embeddings exhibit more analogous roles in the influence propagation process, resulting in community partitions that are more semantically interpretable and aligned with real-world propagation mechanisms.
To account for the impact of community structure on node importance, in this paper, we employ a widely used community detection method, K-Means, to cluster local influence features in multi-layer social networks based on node local influence embeddings $h_G$.

Subsequently, we learn the node community-level embeddings based on the results of community detection, since individuals are more inclined to share information with others who share similar traits according to social identity theory.
Larger communities typically encompass more nodes and exhibit greater connectivity, providing more pathways for information dissemination.
As a result, nodes within such communities are likely to exert influence over a broader set of nodes, either directly or indirectly.
Consequently, the influence range of a node may be correlated with its community size, denoted as $\text{CSize}(v)$, which is formally defined in Equ. (\ref{equ:chap4:11}).
\begin{equation}
  \label{equ:chap4:11}
  \text{CSize}(v)=\frac{1}{|\mathcal{C}(v)|}\sum_{C\in \mathcal{C}(v)}|C|,
\end{equation}
where $\mathcal{C}(v)$ represents the set of all communities to which node $v$ belongs, and $|C|$ denotes the number of nodes in community $C$.
Conversely, a node with a higher number of overlapping communities, denoted as $\text{COverlap}(v)$, indicates multi-membership in multiple communities, suggesting that it participates in a greater number of social groups.
Such nodes are typically regarded as bridges or intermediaries across multiple social groups or communities, rendering them significant for cross-layer propagation in multi-layer networks.
Specifically, the number of overlapping communities $\text{COverlap}(v)$ for a node is formally defined in Equ. (\ref{equ:chap4:12}).
\begin{equation}
  \label{equ:chap4:12}
  \text{COverlap}(v)=|\{\,C\in \mathcal{C}\mid v\in C\,\}|
\end{equation}

$\text{COverlap}(v)$ quantifies the degree to which node $v$ participates in multiple social communities, typically regarded as a critical factor in facilitating cross-community information propagation.
As we mentioned previously, seed sets with highly overlapping influence ranges may result in limited overall influence propagation coverage.
Therefore, in the multi-layer network influence maximization problem, the seed set should be selected to maximize coverage across diverse nodes, ranges, and paths in the networks. 
According to the semantics of the node local influence embeddings $h_G$, seed nodes with proximate embeddings tend to have overlapping influence diffusion ranges and nodes.
Hence, choosing nodes that are distant in the embedding space typically ensures coverage of distinct local propagation structures and more independent influence regions.
To formalize this differentiation based on node embeddings, the Euclidean distance between nodes based on local influence embeddings is calculated as shown in Equ. (\ref{equ:chap4:13}).
\begin{equation}
  \label{equ:chap4:13}
  \text{D}\left( u \right)=  \left \| h_u - h_v \right \|,\  \forall v \in \mathcal{V} \wedge v \neq  u,
\end{equation}
where $\text{D}\left( u \right) \in \mathbb{R}^{n \times d}$.
The final embeddings of nodes are obtained by concatenating the community-level information with the distance embedding vector, as shown in Equ. (\ref{equ:chap4:14}).
\begin{equation}
  \label{equ:chap4:14}
  h_v = \left [\text{CSize} (v) \oplus  \text{COverlap} (v) \oplus  \text{D}\left(  v\right) \right ]
\end{equation}

\subsection{Global Influence Regression Prediction}

\subsubsection{Global Influence Label Generation}
\label{label_gen}

The pseudo-regression task is a supervised learning problem that requires generating continuous influence scores for each node as ground truth labels, which are obtained using the Multi-SIR epidemic model.
Specifically, the Multi-SIR model classifies nodes into three states: Susceptible ($S$), Infected ($I$), and Recovered ($R$).

Initially, all nodes are in the susceptible state ($S$), except for a small set of seed nodes designated as infected ($I$).
Susceptible nodes can be infected by their infected neighbors with a probability $\beta$.
In the Multi-SIR model, once a node is infected in one layer, it has a probability $\theta$ of infecting the corresponding node in other layers, thereby enabling cross-layer influence propagation.
Subsequently, an infected node transitions to the recovered state ($R$) with probability $\gamma$, and the Multi-SIR model assumes that the recovered nodes are immune and cannot be reinfected.
The propagation process is terminated when no further infections occur.
The influence of a seed node is quantified by the total number of nodes in the infected and recovered states at the end of the diffusion process.
In the experiments, the recovery probability $\gamma$ is set to 1, and the infection probability $\beta$ is set to the infection threshold as shown in Equ. (\ref{equ:chap4:15}).
\begin{equation}
  \label{equ:chap4:15}
  \beta_{\text{th}}^{l} = \frac{1}{\bar{d}^{l}}
\end{equation}
where $\beta_{\text{th}}^{l}$ is the critical infection threshold in the $l$-th layer, and $\bar{d}^{l}$ is the average degree of the network in the $l$-th layer.
To mitigate the impact of randomness in simulation results, 1,000 Monte Carlo simulations are conducted. 
The average influence score across these simulations is adopted as the global influence label for each node.

\subsubsection{Influence Score Prediction}

The final embeddings of all nodes, as shown in Equ. (\ref{equ:chap4:14}), serve as inputs to a regressor to predict the influence scores of nodes, which serve as the basis for selecting the seed set.
The predicted influence scores of the regressor are then fitted to the generated global influence labels, and we use the Mean Squared Error (MSE) as the loss function in Equ. (\ref{equ:chap4:16}).
\begin{equation}
  \label{equ:chap4:16}
  \mathcal{L}_{\text{MSE}} = \frac{1}{n}\sum_{i=1}^{n}\big(y_{v_i} - f(h_{v_i})\big)^{2},
\end{equation}
where $f(h_{v_i})$ denotes the predicted influence score of node $v_i$ and $y_{v_i}$ is the generated global influence label.
The nodes are then ranked in descending order based on these predicted influence scores, and the top-$k$ nodes are chosen as the seed set to initiate the diffusion process for influence maximization tasks.

\begin{table}[t]
  \centering
  \caption{The statistics of datasets.}
  \label{tab:chap4:1}
  \resizebox*{\linewidth}{!}{
  \begin{threeparttable}
    \begin{tabular}{cccccc}
      \toprule
       Dataset                    &  Layers        & Nodes             & Edges & $\bar{d}^{l}$ & $\beta_{\text{th}}^{l}$ \\
      \midrule

      \multirow{4}{*}{TailorShop}    & \multirow{4}{*}{4} & \multirow{4}{*}{39}    & 158       & 8.10         & 0.123      \\
                                     &                    &                        & 223       & 11.43        & 0.087      \\
                                     &                    &                        & 76        & 3.89         & 0.257      \\
                                     &                    &                        & 95        & 4.87         & 0.205      \\
      \cline{4-6} 

      \multirow{3}{*}{LazegaLawyers} & \multirow{3}{*}{3} & \multirow{3}{*}{71}    & 717       & 20.19        & 0.049       \\
                                     &                    &                        & 399       & 11.23        & 0.088       \\
                                     &                    &                        & 378       & 10.64        & 0.093       \\
      \cline{4-6} 

      \multirow{3}{*}{CKM}           & \multirow{3}{*}{3} & \multirow{3}{*}{241}   & 449       & 3.72         & 0.269      \\
                                     &                    &                        & 498       & 4.13         & 0.242      \\
                                     &                    &                        & 423       & 3.51         & 0.285      \\
      \cline{4-6} 

      \multirow{3}{*}{SCHOLAT}       & \multirow{3}{*}{3} & \multirow{3}{*}{2,302} & 11,393    & 9.89         & 0.101       \\
                                     &                    &                        & 139,004   & 120.76       & 0.008       \\
                                     &                    &                        & 70,226    & 61.01        & 0.016       \\
      \bottomrule
    \end{tabular}
  \end{threeparttable}
    }
\end{table}

\begin{table*}[t]
    \centering
    \caption{The influence scale $N(k)$ of different methods via three information diffusion models (Multi-SIR, Multi-IC, and Multi-LT) on four real-world datasets.}
    \resizebox*{\linewidth}{!}{
    \begin{threeparttable}
        \begin{tabular}{lccc|ccc|ccc|ccc}
        \toprule
        \multirow{2}{*}{Models} & \multicolumn{3}{c}{TailorShop} & \multicolumn{3}{|c}{LazegaLawyers} & \multicolumn{3}{|c}{CKM} & \multicolumn{3}{|c}{SCHOLAT} \\
        \cmidrule{2-4} \cmidrule{5-7} \cmidrule{8-10} \cmidrule{11-13}
        & Multi-SIR & Multi-IC & Multi-LT & Multi-SIR & Multi-IC & Multi-LT & Multi-SIR & Multi-IC & Multi-LT & Multi-SIR & Multi-IC & Multi-LT \\
        \midrule
        DC & 33.7 & 32.8 & 37.7 & 54.2 & 48.8 & \underline{59.3} & 158.1 & 146.8 & 47.0 & 1255.1 & 1198.9 & 497.7 \\
        CC & 33.7 & 32.6 & 37.7 & 54.6 & 49.1 & \underline{59.3} & 89.1 & 85.3 & 42.9 & 1301.1 & 1243.1 & 1209.8 \\
        BC & 34.1 & 32.9 & 37.7 & \underline{54.9} & 49.6 & 57.9 & 122.4 & 115.1 & 48.1 & 1406.8 & 1343.5 & 1494.4 \\
        ISBC & 33.8 & 32.6 & 37.9 & 52.9 & 45.7 & 12.8 & 100.2 & 94.1 & 35.7 & 1331.8 & 1265.3 & 1225.9 \\
        K-Shell & 32.4 & 30.7 & 31.8 & 52.6 & 45.8 & 20.1 & 89.3 & 83.9 & 37.3 & 1222.6 & 1153.1 & 345.8 \\
        IM-ELPR & 33.6 & 32.1 & 36.4 & 54.8 & 48.2 & 13.1 & \underline{167.2} & 153.7 & 42.3 & 1394.4 & 1321.5 & 360.2 \\
        \midrule
        KSN & 33.3 & 34.2 & 16.0 & 53.5 & 53.0 & 11.7 & 125.2 & 149.4 & 83.9 & 1482.2 & 1530.7 & 246.8 \\
        ISF & 31.8 & 32.8 & 25.2 & 53.2 & 51.3 & 17.8 & 158.4 & 174.7 & 43.3 & 1457.4 & 1484.8 & 1157.7 \\
        CIM & 31.7 & 33.0 & 24.9 & 54.1 & 53.0 & 54.7 & 152.0 & 165.1 & 52.2 & 1376.9 & 1404.2 & 332.4 \\
        DPSOMIM & \underline{34.3} & \underline{34.6} & \underline{37.8} & 54.8 & \underline{53.7} & 21.2 & 162.9 & \underline{178.0} & \underline{85.9} & \underline{1526.2} & \underline{1534.5} & \underline{1538.2} \\
        MIM-Reasoner & 33.1 & 33.9 & 13.7 & 52.8 & 51.4 & 12.1 & 144.1 & 172.3 & 36.3 & 1507.2 & 1526.2 & 421.3 \\
        \midrule
        \textbf{Inf-MDE} & \textbf{34.5} & \textbf{35.1} & \textbf{37.9} & \textbf{62.0} & \textbf{59.6} & \textbf{62.5} & \textbf{173.8} & \textbf{188.4} & \textbf{109.6} & \textbf{1574.6} & \textbf{1621.0} & \textbf{1564.3} \\
	\bottomrule
        \end{tabular}
    \end{threeparttable}
    }
    \label{comparison_result}
\end{table*}

\section{Experimentals}


\subsection{Datasets}

We conduct extensive experiments on four real-world multi-layer social network datasets, including TailorShop~\cite{210-1}, LazegaLawyers~\cite{210-2}, CKM~\cite{210-3}, and SCHOLAT\footnote{\url{https://scholat.com/research/opendata/#multiplex_network}}.
The statistic of datasets are shown in Table \ref{tab:chap4:1}, where $\beta_{\text{th}}^{l} = \frac{1}{\bar{d}}$ and $\bar{d}^{l}$ are the critical infection threshold and the average degree of the $l$-th layer social network. 

\subsection{Evaluation Metrics}

We use two widely used metrics to measure the performance of the seed sets in multi-layer social networks for the influence maximization problem, which are Influence Scale $N(k)$ and Average Distance between Seeds $D_{avg}$.

The influence scale $N(k)$ is used to measure the diffusion capability for a seed set and represents the total number of infected and recovered nodes at the end of the diffusion simulation.
We utilize three commonly used information diffusion models, including Multi-SIR, Multi-IC, and Multi-LT, which enable information propagation across layers.

Average Distance between Seeds $D_{avg}$ is a metric to evaluate the quality of selected seed nodes in a social network.
A higher $D_{avg}$ indicates that the seed nodes are more spread out, potentially covering different communities and improving the spread of influence.
Specifically, $D_{avg}$ is defined as shown in Equ. (\ref{equ:chap4:17}).
\begin{equation}
    \label{equ:chap4:17}
    D_{avg} = \frac{\sum_{u, v \in \mathcal{S}}^{k}D_{u v}}{\binom{k}{2}},
\end{equation}
where $\mathcal{S}$ is the seed set, $D_{u v}$ is the path length between any two nodes $u$ and $v$, and $k$ is the size of seed set.

\subsection{Baselines}

To demonstrate the effectiveness of our proposed model Inf-MDE, we utilize two types of IM methods for performance comparison, including for single-layer networks (DC~\cite{211-2}, CC~\cite{211-5}, BC~\cite{211-3}, ISBC~\cite{211-4}, K-Shell~\cite{211-6}, and IM-ELPR~\cite{210-8}) and for multi-layer networks (KSN~\cite{210-7}, ISF~\cite{210-7}, CIM~\cite{211-1}, DPSOMIM~\cite{210-5}, and MIM-Reasoner~\cite{114-10}).

Our proposed model Inf-MDE is implemented with Python version 3.10 and PyTorch version 2.1.0.
All the experiments are conducted in a Linux Ubuntu 18.04 environment with 256GB of memory and run on two 32GB Nvidia GPUs (Tesla V100 SXM2) with CUDA version 11.8.

\subsection{Experimental Results and Analysis}

\subsubsection{Performance Comparison}

\begin{table}[t]
    \centering
    \caption{The average distance between seeds $D_{\text{avg}}$ of each models across different datasets.}
    \resizebox*{\linewidth}{!}{
    \begin{threeparttable}
        \begin{tabular}{lcccc}
        \toprule
        Models    & TailorShop & LazegaLawyers & CKM   & SCHOLAT \\
        \midrule
        DC           & 1.11              & 1.09                 & 3.09          & 1.08            \\
      CC           & 1.17              & 1.10                 & 1.95          & 1.48            \\
      BC           & 1.22              & 1.16                 & 2.51          & 1.87            \\
      ISBC         & 1.40              & 1.58                 & 2.52          & 1.84            \\
      K-Shell      & 1.09              & 1.24                 & 2.18          & 1.03            \\
      IM-ELPR      & 1.53              & \textbf{1.89}        & 3.97          & 2.31            \\
      \midrule[0.5pt]
      KSN          & 1.71              & 1.73                 & 4.07          & 2.11            \\
      ISF          & 1.20              & 1.33                 & 4.16          & 1.90            \\
      CIM          & 1.11              & 1.02                 & 3.29          & 1.02            \\
      DPSOMIM      & 1.51              & 1.62                 & 3.53          & 2.29            \\
      MIM-Reasoner & \underline{1.77}              & 1.58                 & \underline{4.27}          & \underline{2.71 }           \\
      Inf-MDE      & \textbf{1.78}     & \underline{1.83}     & \textbf{4.32} & \textbf{2.89}   \\
	\bottomrule
        \end{tabular}
    \end{threeparttable}
    }
    \label{tab:chap4:5}
\end{table}

\begin{figure*}[t]
  \centering
  \includegraphics[width=1\textwidth]{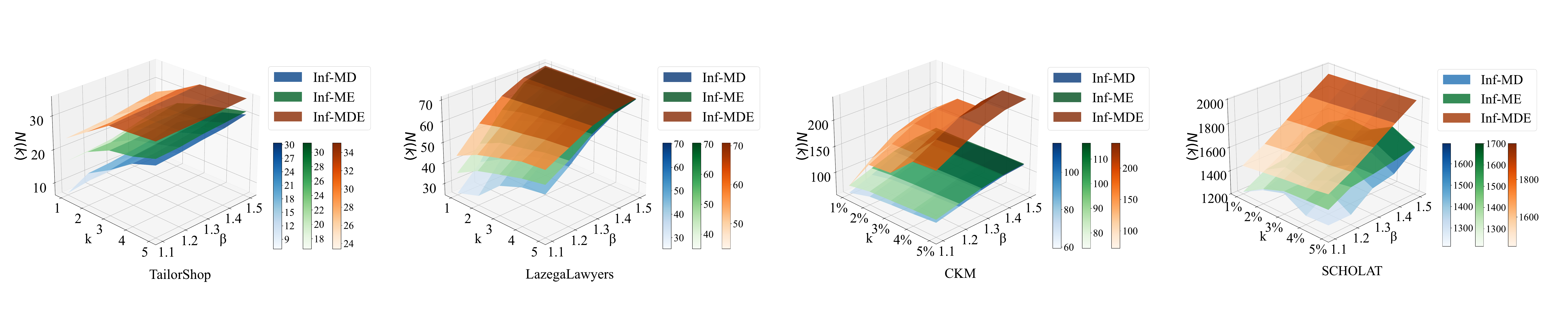}
  \caption{The ablation experimental results of influence scale $N(k)$ via the Multi-SIR model across four datasets.}
  \label{fig:chap4:7}
\end{figure*}

In this section, we demonstrate the overall performance of Inf-MDE and the baselines across four real-world datasets.
For fairly comparisons, we set the size of seed sets $k$ as 10 on TailorShop and LazegaLawyers datasets, and set $k$ as $10\%$ of the number of total nodes on CKM and SCHOLAT datasets.

Table \ref{comparison_result} demonstrates the experimental results of different methods on influence scale $N(k)$ via three information diffusion models, i.e., Multi-SIR, Multi-IC, and Multi-LT, where the highest value of each metric is bold while the second highest value is underlined.
It can be seen that our proposed model Inf-MDE achieves the highest influence scale $N(k)$ across all datasets using different information diffusion models, illustrating the effectiveness for influence maximization tasks.
Traditional centrality-based methods for single-layer social networks, such as DC, CC, and BC, generally underperform in multi-layer scenarios, which indicates that the importance of cross-layer seed nodes is overlooked by single-layer centrality methods.

The influence maximization models for multi-layer social networks, such as Inf-MDE and DPSOMIM, outperform other baselines across four datasets, particularly in the large-scale social network SCHOLAT.
However, even in a sparse network like TailorShop, Inf-MDE maintains its advantage, demonstrating the stability of our proposed model.
This is attributed to the capability of Inf-MDE to incorporate local influence feature embeddings and learn global influence features through regression fitting.
Furthermore, unlike traditional random walk-based subgraphs, Inf-MDE captures the actual influence range of nodes in multi-layer networks by simulating the information diffusion process.

In Table \ref{comparison_result}, we can observe that as the node scale of datasets increases, all models exhibit a gradual rise in influence scale $N(k)$, and our proposed model Inf-MDE exhibits a stronger advantage in datasets with larger node scales.
For example, in the SCHOLAT dataset, the influence scale $N(k)$ of Inf-MDE significantly outperforms other models designed for multi-layer networks, such as DPSOMIM, MIM-Reasoner, and CIM.
In addition, centrality-based methods like DC show significant dataset dependency by using the Multi-LT information diffusion model.
In the sparse network dataset TailorShop, the difference between DC and Inf-MDE is minimal due to the dominance of local influence in simpler structures.
In contrast, in the complex network dataset SCHOLAT, the gap between DC and Inf-MDE widens as the node size grows, underscoring the necessity of learning weights for global influence in multi-layer networks.
This observation aligns with expectations under the Multi-LT model, where seed nodes should maximize mutually exclusive local influence while exhibiting high global influence.

For another evaluation metric, Table \ref{tab:chap4:5} demonstrates the average distance between seed nodes $D_{\text{avg}}$ of different models across four real-world datasets.
It can be seen that our proposed model Inf-MDE achieves the best or second-best results on $D_{\text{avg}}$.
Specifically, Inf-MDE significantly outperforms other baselines on the TailorShop, CKM, and SCHOLAT datasets, while on the LazegaLawyers dataset, the $D_{\text{avg}}$ is slightly lower than IM-ELPR and achieves the second-best result.

\subsubsection{Ablation Experiments}

\begin{table}[t]
    \centering
    \caption{The ablation experiments on average distance between seeds $D_{\text{avg}}$ across four datasets.}
    \resizebox*{\linewidth}{!}{
    \begin{threeparttable}
        \begin{tabular}{lcccc}
        \toprule
        Models    & TailorShop & LazegaLawyers & CKM   & SCHOLAT \\
        \midrule
        Inf-MD    & 1.67              & 1.73                 & 3.34          & 2.67            \\
      Inf-ME    & 1.64              & 1.13                 & 2.62          & 1.04            \\
      Inf-MDE   & \textbf{1.78}     & \textbf{1.81}        & \textbf{4.32} & \textbf{2.89}   \\
	\bottomrule
        \end{tabular}
    \end{threeparttable}
    }
    \label{tab:chap4:6}
\end{table}

In this section, the ablation experiments are conducted to demonstrate the effectiveness of two key components of the proposed model Inf-MDE.
Specifically, we conduct the experiments across four datasets using two ablation models, which are Inf-MD (removing the local influence feature embedding) and Inf-ME (removing the differentiated graph embedding).
Due to the space limitation of this paper, we only demonstrate the results of influence scale $N(k)$ via the information diffusion model Multi-SIR, while employing other information diffusion models (Multi-IC or Multi-LT) can obtain similar results.

The results of ablation experiments are shown in Figure \ref{fig:chap4:7}, where the x-axis represents the sizes of seed sets $k$, the y-axis represents the infection probability $\beta$, and the z-axis and color intensity indicate the results of influence scale $N(k)$. 
The Inf-MDE model, depicted by the green line surface, demonstrates clear superiority compared with the other two ablation models.
In small-scale datasets like LazegaLawyers, high infection rates and larger seed nodes diminish individual node differences, leading to similar performance among the Inf-MDE, Inf-MD, and Inf-ME.
This suggests that in small-scale networks with a large number of seed nodes, despite differences in seed node influence, the propagation outcomes tend to converge.
Conversely, in large-scale networks like SCHOLAT, the Inf-MDE model effectively utilizes local influence features, better capturing node differences and achieving superior performance.

The results of another group of ablation experiments on the average distance between seeds $D_{\text{avg}}$ are shown in Table \ref{tab:chap4:6}, where Inf-MDE still outperforms the other two ablation models.
The reason is that the ego-inf-subgraph constructed by simulating the information diffusion process ensures community division reflects structural similarity in diffusion rather than mere attribute homogeneity.
Furthermore, the differentiated processing of node embedding distance $\mathcal{L}\left( u \right)$ effectively minimizes the overlap of seed sets.

\subsubsection{Sensitivity Analysis of Hyperparameter $step$}
\label{hyper}

During model training, the maximum number of steps for subgraph sampling $step$ is a critical parameter.
A low value of $step$ hinders accurate simulation of the diffusion process, leading to suboptimal feature embeddings, while a large value of $step$ increases computational overhead and causes model over-smoothing.
Table \ref{tab:chap4:7} indicates the effect of $step$ on the influence scale $N(k)$ and average distance between seeds $D_{\text{avg}}$ of Inf-MDE via the Multi-SIR model, using the infection threshold as the criterion and the sizes of seed sets fixed at 5\% of total nodes across four datasets. 

\begin{table}[t]
    \centering
    \caption{The experimental results of Inf-MDE when varying the subgraph sampling step $step$.}
    \resizebox*{\linewidth}{!}{
    \begin{threeparttable}
        \begin{tabular}{lcc|cc|cc|cc}
        \toprule
        \multirow{2}{*}{$step$} & \multicolumn{2}{c}{TailorShop} & \multicolumn{2}{|c}{LazegaLawyers} & \multicolumn{2}{|c}{CKM} & \multicolumn{2}{|c}{SCHOLAT} \\
        \cmidrule{2-3} \cmidrule{4-5} \cmidrule{6-7} \cmidrule{8-9}
        & $N(k)$ & $D_{\text{avg}}$ & $N(k)$ & $D_{\text{avg}}$ & $N(k)$ & $D_{\text{avg}}$ & $N(k)$ & $D_{\text{avg}}$ \\
        \midrule
        0 & 15.76 & 1.17 & 42.67 & 1.13 & 84.13 & 2.62 & 1277.61 & 1.04 \\
        1 & 22.21 & 1.46 & 39.24 & 1.48 & 78.35 & 2.77 & 1077.75 & 1.06 \\
        2 & 22.66 & 1.51 & 40.47 & 1.55 & 79.07 & 3.20 & 1074.41 & 1.55 \\
        5 & \textbf{29.43} & \textbf{1.78} & \textbf{54.75} & \textbf{1.81} & \textbf{182.04} & \textbf{4.32} & 1536.59 & 2.89 \\
        10 & 27.97 & 1.77 & 47.09 & 1.77 & 168.26 & 3.73 & \textbf{1670.10} & \textbf{3.11} \\
	\bottomrule
        \end{tabular}
    \end{threeparttable}
    }
    \label{tab:chap4:7}
\end{table}

As shown in Table \ref{tab:chap4:7}, a small $step$ (e.g., $step=[0,2]$) limits the model to capture the local influence feature between nodes, resulting in a suboptimal $N(k)$ and $D_{\text{avg}}$.
Conversely, a large $step$ (e.g., $step=10$) causes node embeddings to overly aggregate features from distant nodes, blurring local diffusion features and degrading the model performance, especially in some small-scale social networks, such as TailorShop, LazegaLawyers, and CKM.
When the value of $step$ increases, it may lead to feature redundancy and similar performance (e.g., $step \in [5,10]$), necessitating dynamic adjustment based on the network scale.

To further demonstrate the effectiveness of the proposed model Inf-MDE, we conduct two additional groups of sensitivity experiments, varying the size of seed sets $k$ and the infection probability $\beta$ of the diffusion model.
The results and analysis are provided in the appendix.

\section{Conclusion}

In this paper, we propose the Inf-MDE model, a multi-layer influence maximization model based on differentiated graph embeddings for influence maximization tasks on multi-layer social networks.
By combining local influence embeddings with global influence embeddings, the proposed model captures structural information on inter-layer and intra-layer influence diffusion in multi-layer social networks, which enables the model to identify seed nodes effectively.
We demonstrate an advantage over state-of-the-art models on four public datasets and conduct a case study on a proprietary dataset. This case study includes an effect analysis in three areas; details are provided in the Appendix.

\bibliographystyle{aaai2026}
\bibliography{aaai2026}

\appendix

\section{Appendix}

\subsection{Definition of Influence Maximization}

Given a social network represented as an undirected graph \(\mathcal{G} = (\mathcal{V}, \mathcal{E})\), where the node set \(\mathcal{V} = \{v_0, v_1, \ldots, v_n\}\) and each node \(v\) represents a user. The social relationship between user \(v_i\) and \(v_j\) is indicated by the edge \(e = (v_i, v_j)\), which means information can be transmitted from user \(v_j\) to user \(v_i\). The edge set \(\mathcal{E} = \{e_0, e_1, \ldots, e_m\}\) describes all the connections in the social network. Given an information propagation model \(\mathcal{M}\), which can generally be an Independent Cascade model, Linear Threshold model, or Epidemic Propagation model, etc., and a budget \(k\), the goal of the Influence Maximization problem is to select a seed node set \(S \subseteq \mathcal{V}\) of size \(k\) such that the expected number of activated nodes \(\sigma(S)\) is maximized when the influence propagation process ends. Mathematically, this problem can be described by Equ. (\ref{equ:IM}):

\begin{equation}
\label{equ:IM}
S^* = \mathop{\arg\max}\limits_{S \subseteq \mathcal{V}, |S| = k, \mathcal{G}, \mathcal{M}} \sigma(S),
\end{equation}
where \(\sigma(S)\) represents the influence coverage range of the seed set \(S\) on the social network \(\mathcal{G}\) under the given information propagation model \(\mathcal{M}\), computed via Monte Carlo simulation.

\subsection{Related Work}

\subsubsection{Influence Maximization}
As a traditional algorithm, the influence maximization method based on node centrality measures the approximate influence of all nodes in the network and selects the top-k nodes as the seed set, and the key is to determine the appropriate ranking indicators. Common local measures such as degree centrality \cite{2024Normalized,2022MDER} determine influence based on the number of neighbors, but ignore the location of nodes, making it difficult to identify bridge nodes that connect multiple dense communities (i.e., few neighbors but play a key role) \cite{2022LFIC}. Measures based on the global structure include: betweenness centrality \cite{211-2} (i.e., a measure of how often a node acts as a bridge on the shortest path), tight centrality \cite{211-5} (i.e., a measure of the reciprocal of the average proximity of a node to other reachable nodes), eigenvector centrality \cite{2021Blind} (i.e., considering neighbor importance differences and treating influence as a weighted sum of neighbor influences), and similar Katz centrality \cite{2024TATKC}, as well as K-shell decomposition that identifies influence by node topological location (i.e., from edge to core) \cite{211-6}. Although these indicators can reflect the topological characteristics of nodes, their adaptability and generalization ability are limited \cite{2023IMSurvey}, and they are not suitable for large-scale social networks.

Deep learning, especially graph embedding techniques for graph neural networks (GNNs), is a potential solution to the Influence Maximization (IM) problem. The relevant methods include: treating node influence prediction as a classification task (e.g., DeepInf \cite{2018DeepInf}), binary classification to distinguish between high and low influence nodes (e.g., InfGCN \cite{2020InfGCN}), predicting influence value for regression problems (e.g., RCNN \cite{2020RCNN}), learning embedding vectors by combining community perception and skip-gram (e.g., DeepIM \cite{1219-2}), extracting high-order embedding and screening nodes based on meta-path random walk (e.g., MAHE-IM \cite{2022MAHE_IM}), and the application of deep reinforcement learning and probabilistic graph models for cross-layer budget allocation and scheme decomposition (e.g., MIM-Reasoner \cite{2024mim_reasoner}). However, these methods generally have the problems of insufficient feature extraction of interlayer propagation heterogeneity and intralayer diffusion mechanism, and there may be representation bias between their embedded spatial semantics and real propagation dynamics, which seriously restricts the applicability of the algorithm in real social networks.

\subsubsection{Influence Maximization In Multi-layer Networks}
Although the problem of influence maximization in multi-layer social networks is still in its early stages, some research has recently begun to investigate the influence maximization problem in multi-layer social networks \cite{114-1,114-5}.
However, existing approaches based on multi-layer social networks simply sum up the identification results of each layer \cite{114-4}.
These methods treat multi-layer social networks as a few isolated single-layer social networks, ignoring the effects of inter-layer relationships and cross-layer propagation mechanisms that significantly impact the overall spread of information, which biases the analysis of the influence of nodes.
Even if we consider modeling on multi-layer social networks and taking into account the inter-layer relationships of multi-layer social networks, the existing works \cite{114-2,114-8} are based on the traditional node centrality metrics, such as degree centrality, betweenness centrality, and k-shell decomposition.
However, in multi-layer social networks, the influence of nodes is often heterogeneous, and relationships at different layers may have different propagation mechanisms and weights.
Traditional centrality metrics are unable to capture this heterogeneity in cross-layer propagation, i.e., the dynamic performance of nodes in different social scenarios, making it difficult to capture the rich topological information hidden in multi-layer social networks and to obtain more detailed local influence.
To address these problems, researchers have recently attempted to incorporate community structures into the influence maximization framework for multi-layer networks, which refers to a group of individuals in a social network, and has a significant impact on information diffusion due to the fact that their interactions within the community are more frequent than their interactions with individuals outside the community \cite{114-9}.
However, some existing studies have pointed out that it is unreasonable to use the same community structure in all social network layers \cite{114-7}, as the connectivity patterns vary from layer to layer in a multi-layer social network, leading to different community formations in each layer\cite{114-6}.
However, its work after considering the heterogeneity of community structure ends up measuring only node centrality metrics.
In contrast, GNN is able to utilize the topology of the network and the attribute information of the nodes to better capture the position and semantic features of the nodes in the network by learning the high-dimensional embedding representation of the nodes.
The community detection based on node embeddings is not only more accurate, but also can extract community-level feature information, providing richer propagation patterns and structural information.

Deep learning-based approaches to influence maximization have shown strong capabilities in capturing global influence \cite{1219-2}, current studies on influence maximization in multi-layer social networks \cite{114-10} still lack effective modeling of inter-layer diffusion heterogeneity and intra-layer propagation mechanisms, as well as the characterization bias between the semantics of the node embedding space and the characteristics of the propagation dynamics.
That is, the node embedding spaces only represent the static topological similarity without explicitly modeling intra-layer and inter-layer diffusion path dependencies, which leads to a semantic characterization bias between the generated embeddings and the real influence propagation patterns.

\subsection{The Pseudo-Code of Ego-Inf-Subgraph Extraction}

In Inf-MDE, we propose an ego-inf-subgraph sampling method to capture the topology information and influence propagation feature between nodes, where the pseudo-code is shown in Algorithm \ref{alg:chap4:1}.
Since the subgraph sampling process has been detailed in the main text, it will not be repeated here.

\begin{algorithm}[t]
\renewcommand{\algorithmicrequire}{\textbf{Input:}}
\renewcommand{\algorithmicensure}{\textbf{Output:}}
  \caption{The Process of Ego-Inf-Subgraph Extraction}
  \label{alg:chap4:1}
    \begin{algorithmic}[1]
      \REQUIRE multi-layer social network $\mathcal{G}=\{\mathcal{G}_1,\dots,\mathcal{G}_l\}$, inter-layer propagation probability $p^i$, the node set $\mathcal{V}$, cross-layer propagation probability $\theta^i$
      \ENSURE the set of ego-inf-subgraph $\mathcal{F}$
      \FOR{each node $v \in \mathcal{V}$}
      \STATE Initialize activated set $\mathcal{A}_v = \{v\}$
      \FOR{$t=1$ to $step$}
      \STATE Initialize a new activated set $A_{\text{new}} = \emptyset$
      \FOR{each $\mathcal{G}_i \in \mathcal{G}$}
      \FOR{all activated nodes $v$}
      \FOR{each unactivated neighbor $v_j$}
      \IF{successfully activated $v_j$ with probability $p^i$}
      \STATE $A_{\text{new}} \leftarrow A_{\text{new}} \cup \{v_j\}$
      \ENDIF
      \ENDFOR
      \ENDFOR
      \ENDFOR
      \IF{$A_{\text{new}} = \emptyset$}
      \STATE Break
      \ENDIF
      \STATE Update activation set $\mathcal{A}_v \leftarrow \mathcal{A}_v \cup A_{\text{new}}$
      \ENDFOR
      \STATE $\mathcal{F} \leftarrow \mathcal{F} \cup \mathcal{G}_{\text{ego-inf}}(v)$ where $\mathcal{G}_{\text{ego-inf}}(v)$ is the ego-inf-subgraph of $v$ based on $\mathcal{A}_v$, containing the activated nodes and the edges between them
      \ENDFOR
      \RETURN $\mathcal{F}$
    \end{algorithmic}
\end{algorithm}

\subsection{Description of Datasets}

\begin{itemize}[leftmargin=*]
\item TailorShop: 
This dataset consists of the social interaction patterns of workers at a tailor shop in Northern Rhodesia over a ten-month observation period.
The network layers represent two distinct types of interactions: work-related collaboration and friendship/social-emotional exchanges.
\item LazegaLawyers:
The LazegaLawyers dataset derives from a network study of a law firm in the northeastern United States, spanning social relationships among 71 lawyers from 1988 to 1991.
The dataset encompasses three network layers: close coworker collaboration, professional advice exchange, and friendship.
\item CKM: 
CKM dataset examines the diffusion of medical innovations within social networks and investigates how varying social network structures influence the adoption of tetracycline among physicians.
It constructs a three-layer social network comprising medical advice consultations, clinical discussions, and close social contacts.
\item SCHOLAT: 
The SCHOLAT dataset provides multiple relations and information between users on academic social networks.
It includes three layers: friendship among users, team relationships within the same group, and connections among users who share a common course.
\end{itemize}

\subsection{Details of Baselines}

\begin{itemize}[leftmargin=*]

\item DC: 
DC mainly ranks the nodes according to their degree, and the top-$k$ highest-ranked nodes are selected as the seed set.

\item CC: 
Closeness Centrality (CC) computes the reciprocal value of the shortest path distances between a node and all other reachable nodes, and then normalizes and aggregates to assess how centrally located a node is within the network.

\item BC: 
BC quantifies the importance of a node by calculating how frequently it appears on the shortest paths between all pairs of nodes, referred to as betweenness.

\item ISBC: 
ISBC is an IM method based on isolating-betweenness centrality.
It is used to measure node influence by fusing betweenness centrality and isolating-betweenness centrality.

\item K-Shell: 
K-Shell decomposition iteratively removes nodes with low degrees to partition the network into distinct core layers, and uses the shell value of the node as an important indicator of its centrality.
It provides a more rich representation of a node's position within the core structure of the network.

\item IM-ELPR:
IM-ELPR is an IM method that integrates the extended H-index, label propagation, and a relationship matrix.
It initially ranks nodes using the Extended H-index centrality measure and identifies the communities within the network via label propagation technique.
Subsequently, smaller communities are consolidated into larger ones using a relationship matrix, from which top-$k$ nodes are selected as the final seed set.

\item KSN: 
KSN is a knapsack-based IM method, which first decomposes a multi-layer network into a single layer, and addresses influence maximization problems on each single layer independently.

\item ISF: 
ISF is a greedy strategy-based influence maximization method for multi-layer networks in multi-layer influence maximization task.

\item CIM: 
CIM is a clique detection-based method for multi-layer influence maximization.
It identifies cliques in the network to locate nodes with potentially high diffusion capabilities. 

\item DPSOMIM: 
DPSOMIM is a discrete particle swarm optimization-based method for influence maximization.
It selects seed nodes by leveraging random connectivity centrality measures to assess node importance from a connectivity perspective.
It utilizes the multi-layer expected diffusion value as the fitness function and incorporates neighborhood optimization to enhance the model's convergence.

\item MIM-Reasoner: 
MIM-Reasoner is an influence maximization method based on deep reinforcement learning and probabilistic graphical models.
It decomposes multi-layer networks into a single-layer network and uses reinforcement learning to select seed nodes.
It estimates cross-layer influence via probabilistic graphical models and adopts a layer-wise training strategy to simulate the overall diffusion process.

\end{itemize}

\begin{figure*}[t]
    \centering
    \subfloat[Multi-SIR model]{
    \label{seed_size:sir}
    \includegraphics[width=0.95\textwidth]{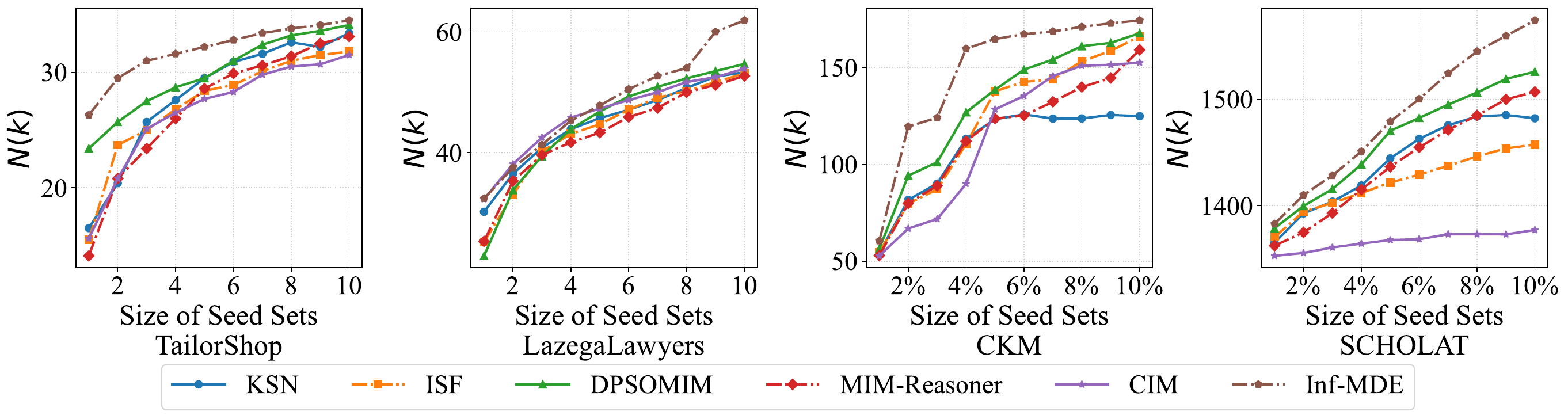}
    } 
    \quad 
    \subfloat[Multi-IC model]{
    \label{seed_size:ic}
    \includegraphics[width=0.95\textwidth]{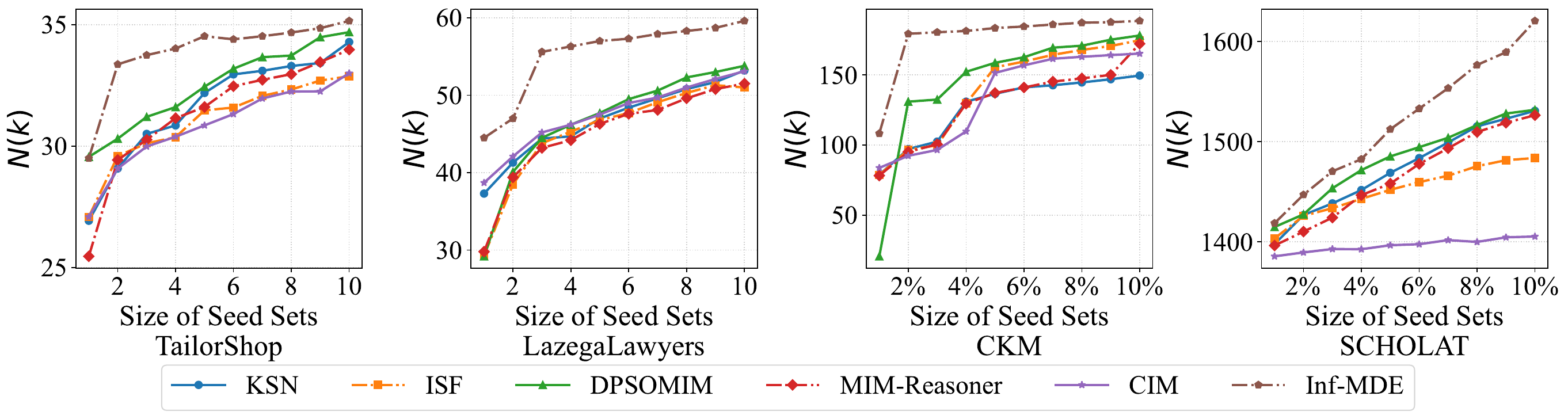}
    }
    \quad
    \subfloat[Multi-LT model]{
    \label{seed_size:lt}
    \includegraphics[width=0.95\textwidth]{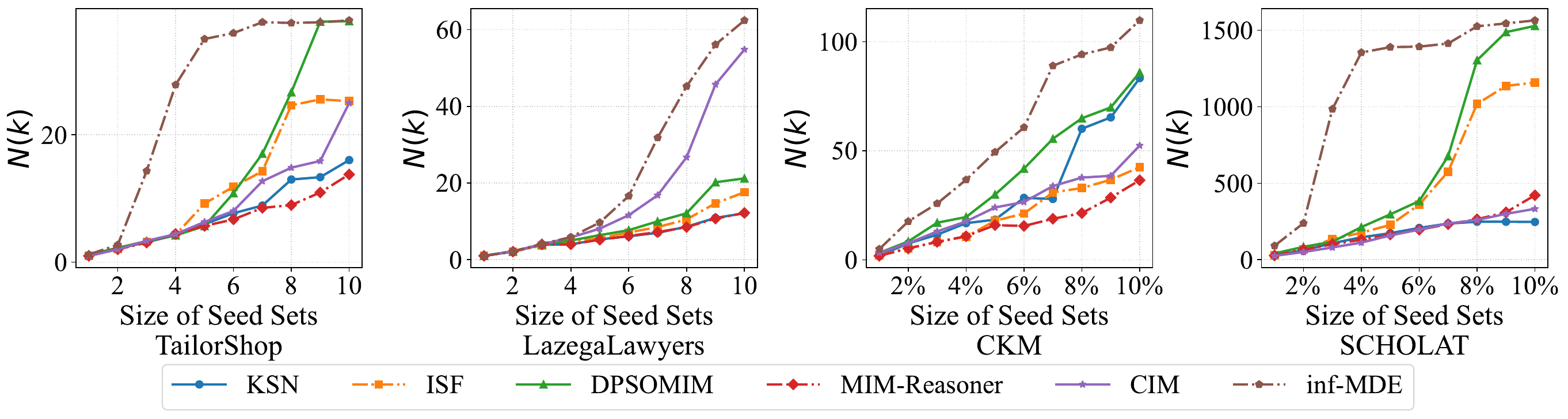}
    }
    \caption{The experimental results with different sizes of seed sets by using Multi-SIR, Multi-IC, and Multi-LT models.}
    \label{seed_size}
\end{figure*}

\begin{figure*}[t]
  \centering
  \includegraphics[width=1\textwidth]{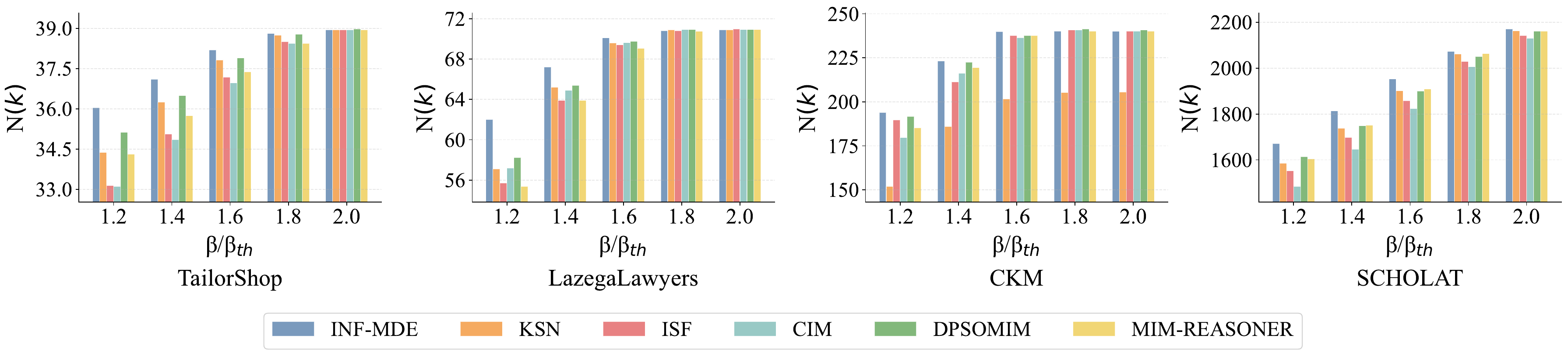}
  \caption{The experimental results with different infection probabilities $\beta$ via Multi-SIR model.}
  \label{beta_result}
\end{figure*}

\subsection{Performance Comparison with Different Sizes of Seed Sets $k$}

To further demonstrate the effectiveness of the proposed model Inf-MDE, we conduct another group of comparative experiments by varying the size of seed sets $k$.
In this paper, we only illustrate the experimental results on the metric influence scale $N(k)$, while we can get similar results on another metric, the average distance between seeds $D_{avg}$.
For the TailorShop and LazegaLawyers datasets, we vary $k$ with values from 1 to 10, while for the CKM and SCHOLAT datasets, $k$ is varied from 1\% to 10\% of the total number of nodes.

As shown in Figure \ref{seed_size}, our proposed model Inf-MDE achieves the highest or near-highest influence scale $N(k)$ across all datasets by varying the size of seed sets, which demonstrates its effectiveness in addressing the influence maximization problem.
For small-scale datasets like TailorShop and LazegaLawyers, when the size of seed sets is small, the performance of Inf-MDE is comparable to some baseline models, particularly using the Multi-LT model.
The reason is that Multi-LT is a linear threshold-based model, and its influence propagation mechanism in multi-layer social networks differs significantly from the Multi-SIR and Multi-IC, which is primarily according to the cumulative influence of neighbors rather than independent probabilities.
However, as the size of seed sets $k$ increases, the models designed for multi-layer networks, such as Inf-MDE and DPSOMIM, exhibit a clear performance advantage.
This validates their ability to effectively tackle the influence maximization problem in multilayer social networks, especially in large-scale networks.

\subsection{Performance Comparison with Different Infection Probabilities $\beta$}

In the Multi-SIR model, $\beta$ is a parameter that controls the probability of infection spread, with higher $\beta$ values indicating a greater likelihood of propagation.
In this group of experiments, we select top-$k$ seed nodes using Inf-MDE and conduct influence propagation via the Multi-SIR model by varying the infection probability $\beta$.
For the TailorShop and LazegaLawyers datasets, $k$ is set as 10, while for the CKM and SCHOLAT datasets, $k$ is set as 10\% of the total number of nodes.
We only illustrate the experimental results on the metric $N(k)$, while we can get similar results on another metric $D_{avg}$.

The experimental results are shown in Figure \ref{beta_result}, where in the x-axis, $\beta_{\text{th}}$ is the constant infection threshold.
As shown in Figure \ref{beta_result}, as the infection probability ratio $\beta / \beta_{\text{th}}$ increases, all the models eventually achieve the same or similar influence scale $N(K)$, which we call the peak value of influence propagation.
However, our proposed model Inf-MDE can reach this peak value faster than other baselines. 
This is because Inf-MDE effectively considers overlapping communities ($\text{COverlap}(v)$) and community size ($\text{CSize} (v)$) and adjusts their weights by using learnable parameters, enabling the model to identify cross-layer bridging nodes.

\begin{figure*}[t]
  \centering
  \includegraphics[angle=90, width=0.55\textwidth]{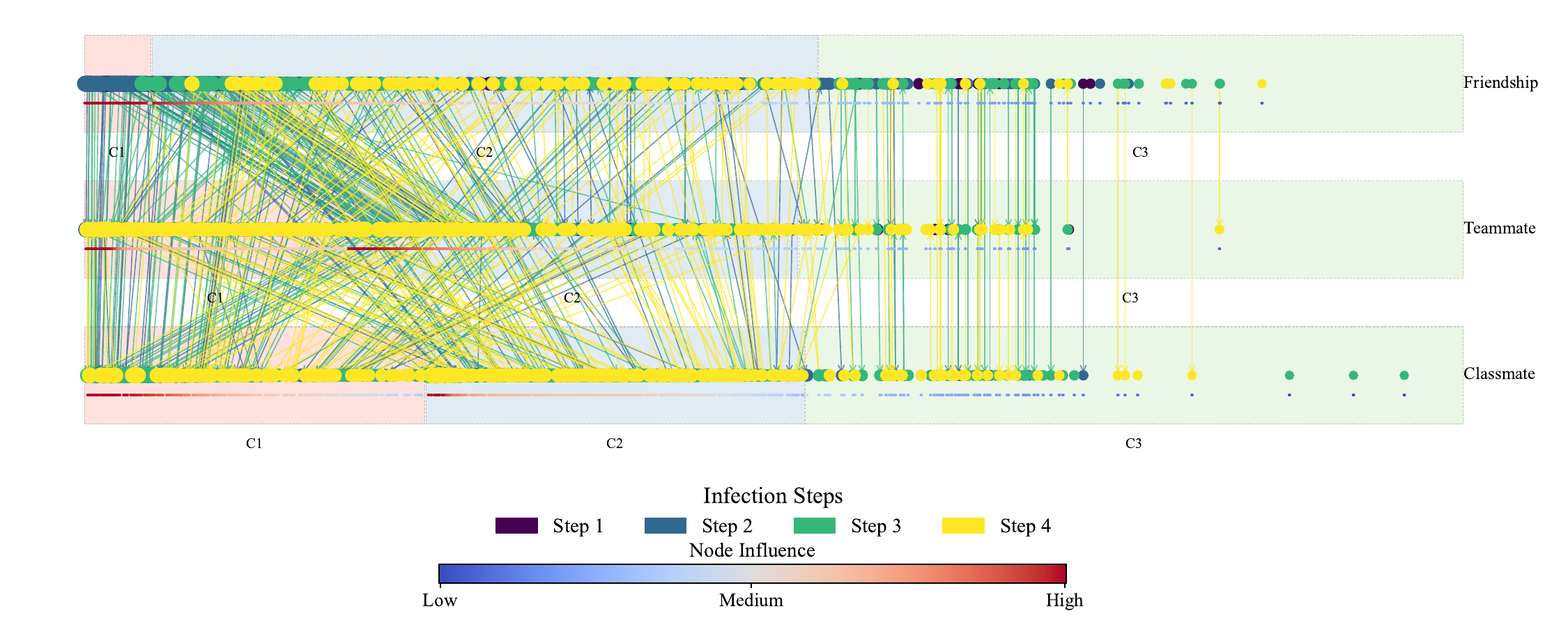}
  \caption{A case study of influence diffusion process in SCHOLAT dataset.}
  \label{case_study}
\end{figure*}

\subsection{Case Study}

In this section, we propose a case study to demonstrate the influence diffusion process in the SCHOLAT dataset, which is a three-layer social network for teaching and research collaboration with three types of relationships between nodes (friendship, teammate, and classmate).
The influence propagation process across three layers is illustrated in Figure \ref{case_study}.
Following the pipeline of Inf-MDE, we get the local influence feature embeddings and conduct community detection in each layer with distinct background colors indicating different communities. 
To enhance the readability of the figure, we only depict the first four steps of the propagation process via the Multi-IC model.
Nodes and edges are colored differently to indicate that they were infected at a specific step, as shown in the legends of Figure \ref{case_study}. 
The dots below the nodes represent their influence scores, as predicted by Inf-MDE, with colors closer to red indicating greater influence scores and those closer to blue indicating lesser influence scores.

In this case study, we obtain the following findings.

\begin{itemize}[leftmargin=*]
\item \textbf{Cross-layer influence propagation predominantly involves high-influence nodes.}
High-influence nodes exhibit significantly higher edge density in cross-layer propagation than low-influence nodes.
This aligns with real-world scenarios, where high-influence users typically wield substantial influence across multiple social networks, and their opinions or comments are more likely to propagate across these networks.

\item \textbf{Community clustering reveals influence biases in cross-layer influence propagation.}
In this case study, we divide the social network of each layer in the SCHOLAT dataset into three communities, denoted as C1, C2, and C3.
We can see that the cross-layer influence propagation is much denser in the C1 and C2 communities, reflecting that nodes within these communities have greater influence and are more likely to perform cross-layer propagation.

\item \textbf{The direction of cross-layer influence propagation yields further insights.}
(1) The influence propagation from the friendship layer to the other two layers (teammate and classmate) is denser, suggesting that information within the friendship layer may impact the behaviors in the team and course contexts.
(2) Compared to the classmate layer, the teammate layer shows a stronger tendency to propagate influence toward the friendship layer, indicating that team members are more likely to share information within their friend circles to seek research collaboration or assistance.
(3) Influence propagation from the classmate layer to the other two layers indicates that the teaching behaviors may affect the teachers' or students' behaviors in team or friend circles.
Course-related information may spread to the friend circle through students, for example, when a student takes a course, he/she may recommend it to friends.

\end{itemize}

\end{document}